\documentclass[fleqn,usenatbib]{mnras}

\usepackage{newtxtext,newtxmath}

\usepackage[T1]{fontenc}

\DeclareRobustCommand{\VAN}[3]{#2}
\let\VANthebibliography\thebibliography
\def\thebibliography{\DeclareRobustCommand{\VAN}[3]{##3}\VANthebibliography}

\usepackage{graphicx}
\usepackage{dcolumn}
\usepackage{bm}
\usepackage{bigints}
\usepackage{multirow}

\usepackage{hyperref}
\usepackage[capitalize]{cleveref}
\usepackage[caption=false]{subfig}
\newcommand{\phantomsubfloat}[1]{
    {
        \captionsetup[subfigure]{labelformat=empty}
        \subfloat[][]{#1}
    }%
}

\title[ML search for never-seen GW sources]{A semi-supervised Machine Learning search for never-seen Gravitational-Wave sources}

\author[T. Marianer et al.]{
Tom Marianer,$^{1}$\thanks{E-mail: tommaria@mail.tau.ac.il}
Dovi Poznanski$^{1}$
and J. Xavier Prochaska$^{2}$
\\
$^{1}$School of Physics and Astronomy, Tel-Aviv University, Tel-Aviv 69978, Israel\\
$^{2}$Department of Astronomy \& Astrophysics, UCO/Lick Observatory, University of California, 1156 High Street, Santa Cruz, CA 95064, USA
}

\date{Accepted XXX. Received YYY; in original form ZZZ}

\pubyear{2020}

\begin{document}
\label{firstpage}
\pagerange{\pageref{firstpage}--\pageref{lastpage}}
\maketitle

\begin{abstract}
By now, tens of gravitational-wave (GW) events have been detected by the LIGO and Virgo detectors. These GWs have all been emitted by  compact binary coalescence, for which we have excellent predictive models. However, there might be other sources for which we do not have reliable
models. Some are expected to exist but to be very rare (e.g., supernovae), while others may be totally unanticipated.
So far, no unmodeled sources have been discovered, 
but the lack of models makes the search for such sources much more difficult and less sensitive. We present here a search for unmodeled GW signals using semi-supervised machine learning. We apply deep learning and outlier detection algorithms to labeled spectrograms of GW strain data, and then search for spectrograms with anomalous patterns in public LIGO data. We searched $\sim 13\%$ of the coincident data from the first two observing runs. No candidates of GW signals were detected in the data analyzed. We evaluate the sensitivity of the search using simulated signals, we show that this search can detect spectrograms containing unusual or unexpected GW patterns, and we report the waveforms and amplitudes for which a $50\%$ detection rate is achieved.

\end{abstract}




\section{Introduction}

Since the first observation of the gravitational wave (GW) signal from a binary black hole (BBH) merger in 2015 \citep{abbott2016observation}, tens of additional signals from BBH, binary neutron star (BNS) and neutron star-black hole (NSBH) mergers have been detected during the first three LIGO-Virgo observing runs \citep{abbott2019gwtc,gracedb}. Many of these observations are the first of their kind, such as the first observation of a BBH merger \citep{abbott2016observation}, the first observation of a BBH system with asymmetric masses \citep{Abbott_2020}, and perhaps most importantly, the first multi-messenger observations of a BNS merger in GWs with its electromagnetic signature \citep{Abbott_2017,2019arXiv191205659N}.

The GW waveform at the detector for all these merger events can be accurately modeled. 
Indeed, LIGO relies on searches by matched-filtering a template bank of modeled waveforms with GW detector strain data to discover these merger events \citep{usman2016pycbc,sachdev2019gstlal}. 
In the presence of additive stationary Gaussian noise only, this search method is guaranteed to be optimal by the Neyman-Pearson lemma \citep{neyman1933ix,ofek2018optimal}. All of the events reported by the LIGO-Virgo Collaboration (LVC) make up a very small fraction of the data collected by the detectors ($<100 \mathrm{s}$ out of an order of  $10^7 \mathrm{s}$ of strain data), with additional signals likely 
existing in the data below the sensitivity threshold of the LVC pipelines. This is illustrated by searches conducted by independent groups which report the detection of additional events \citep[e.g.][]{venumadhav2019new}.

These search methods however are suboptimal for unmodeled sources, and may miss them entirely. These could be core collapse supernovae (CCSNe), mergers with parameters ouside the range covered by the template bank used in the search, or other sources of unknown origin. While perhaps unlikely, one should be open to the possibility of the serendipitous discovery of an unexpected source of GWs. For this purpose, the LVC developed and runs the coherent WaveBurst (cWB) pipeline \citep{klimenko2008coherent,klimenko2016method} which searches for excess energy in time-frequency spectrograms coincident in several GW detectors. So far, no unmodeled source has been discovered with cWB, but the pipeline did trigger on over half of the merger signals in the first two observing runs \citep{abbott2019gwtc}. 

We suggest here a method to dig deeper into the noise and extend the range of the search for unmodeled sources. We do this by looking for coincident signals of any sort in the datastreams from multiple detectors, using their time-freqency patterns. These can serve as tell-tale signs that they share the same physical origin. We do so while attempting to minimally influence the search with our preconceived notions as to the sources' characteristics. 

Pattern recognition is a field that machine learning (ML) algorithms excel at, with significant advancements in recent years. ML techniques have been widely used to enhance GW science (see review by \citealt{cuoco2020enhancing}), however there are currently no published ML-based searches for unmodeled sources. As we were finishing the work on this project, \citet{skliris2020real} proposed another promising ML-based method, but they have not yet performed a search for new signals in existing data. 

Our endeavour could be framed as a task for anomaly (or outlier) detection algorithms. Outlier detection algorithms were used in a variety of astronomical applications (see review by \citealt{baron2019machine}). The more common use of such algorithms in astronomy is to search for anomalous objects in large datasets coming from large surveys. For instance, \citet{baron2017weirdest} applied an algorithm based on random forest to galaxy spectra and found multiple interesting sources, and \citet{reis2018detecting} did the same with stellar spectra. \citet{giles2019systematic} detected outliers in Kepler light curves using density-based clustering. \citet{ralph2019radio} used a combination of a self-organizing map and a convolutional autoencoder on radio-astronomy images, and \citet{hocking2018automatic} developed a method comprised of several algorithms and applied it to optical images.

Since we do not have a template for the patterns we are looking for we need to learn what the `normal' patterns are, and then identify spectrograms containing patterns that are different from the `normal' ones. Since there are very few GW events, most of the `normal' patterns consist of background noise, and the task is to identify sources that are rare and not consistent with being noise. In general, outlier detection algorithms model the distribution of the input samples, and use some measure to flag samples that are unlikely to be drawn from the modeled distribution. This can be achieved in a completely unsupervised manner, by using all samples to model the distribution without any prior knowledge of the data. Alternatively, in a semi-supervised setting, one uses knowledge about `normal' samples, inliers, to model the  distribution and to identify outliers. Here, we apply a semi-supervised approach.

In order to model the distribution of `normal' spectrogram patterns in the data we use deep learning, which is the technique of constructing and training deep neural networks. Neural networks are complex non-linear functions with many parameters, constructed by layers of linear combinations followed by non-linear activation functions. They are commonly used in classification tasks, where each training sample has a target label, and they are trained by tuning the parameters to give the best fit between the training samples and their labels. State-of-the-art deep networks consist of tens or hundreds of layers. One of the more successful architectures of connecting between the layers is the convolutional architecture, where the `neurons' in each layer are connected to a small localized subset of the `neurons' in the previous layer, initially proposed for image processing tasks, since pixels in an image have highly localized correlations. Indeed, the most significant achievements in image classification in recent years were made by deep convolutional neural networks (CNN).

Deep networks trained for classification can be used for outlier detection in two ways. The first way is to use the network for feature extraction, usually by truncating the network and extracting one of the layers before the output layer. Then, traditional outlier detection algorithms can be applied to the extracted features. The motivation for using a deep network for feature extraction is that these networks can learn representative features during training, and therefore require little pre-processing and feature engineering (see for instance \citealt{erhan2009visualizing,zeiler2014visualizing,yosinski2015understanding}). The second way is to use the network directly in the context of out-of-distribution (OOD) detection. Deep networks trained for classification generally `view' the world as containing only the categories on which they were trained (referred to as in-distribution), forcing  all inputs they receive into one of them, assuming all inputs are drawn from the same distribution as the training data. A network trained to separate cats from dogs, when presented with the image of a chair, will look for its whiskers. Since these algorithms are now routinely deployed in the real world, sometimes in life-endangering settings (e.g., autnonomous driving), detecting when an input is drawn from outside that distribution is critically important. Hence the advances in recent years in the field of OOD detection by deep networks \citep[e.g., ][]{hendrycks2016baseline,lee2018simple,sastry2019detecting}.

We implement a search by training a deep CNN on the Gravity Spy glitch dataset \citep{zevin2017gravity,bahaadini2018machine}. Glitches are non-Gaussian noise transients that often occur in GW detector data, and this dataset contains spectrograms of many of these glitches, labeled into various glitch categories. 
\citet{george2017deep} use the transfer learning method to fine-tune a pre-trained CNN to classify the Gravity Spy glitches, and demonstrate this can result in state-of-the-art classification accuracy. We use the same method to train our network, and then apply two methods of outlier detection to the trained network. We apply our search to a significant subset of public LIGO data from the first two observing runs \citep{abbott2019open} and review our findings which include no promising candidate source. We test our method via simulations. 

This paper is structured as follows. In \cref{sec:train} we describe how we train a deep CNN on GW spectrograms. In \cref{sec:outlier_detection} we describe outlier detection methods we use to flag candidate spectrograms. In \cref{sec:search} we describe how the CNN and outlier detection methods are used to search for signals in the public LIGO data. The results of this search are presented and discussed in \cref{sec:results}. In \cref{sec:eval} we evaluate the sensitivity of our search using simulated signals, and we conclude in \cref{sec:conlustions}.

\section{Training phase}
\label{sec:train}

\subsection{Generating Gravity Spy spectrograms}
\label{ssec:gen_spec}

In this search we train a deep CNN using the Gravity Spy dataset, described in detail by \citet{zevin2017gravity} and \citet{bahaadini2018machine}.
This dataset consists of spectrograms of GW strain data from the two LIGO detectors, from the first two LIGO observing runs, labeled into 22 different classes. 20 of these classes are different glitch classes (various noise transients that create particular patterns in the spectrograms) and the final two are the `No Glitch' class -- spectrograms of background detector noise -- and `Chirp' -- spectrograms of simulated BBH mergers.

For the sake of consistency with later stages of our pipeline, we do not use the existing spectrograms from the Gravity Spy dataset directly, but generate spectrograms from the raw strain data, using the time-stamps of the events in the Gravity Spy dataset, and label them using the Gravity Spy labels. These spectrograms are generated using the same process used to generate the unlabeled spectrograms described in \cref{ssec:search_gen}. For each event in the Gravity Spy dataset this process is applied to a segment of $64 \mathrm{s}$ of strain data around the event time-stamp, and the 2 second spectrogram containing the event is saved.

We note that in the Gravity Spy dataset, for each event there are four different spectrograms, of different lengths: $0.5 \mathrm{s}, 1 \mathrm{s}, 2 \mathrm{s}, 4 \mathrm{s}$. At this stage, we limit ourselves to spectrograms of $2 \mathrm{s}$, following \citet{george2017deep} who found this duration is well suited for most glitches.

\subsection{Splitting and augmenting the dataset}
\label{ssec:split}

The dataset generated consists of 5928 spectrograms labeled into 20 different classes. The reason this is smaller than the Gravity Spy dataset is that for some of the events in Gravity Spy the raw strain data were not publicly available, in particular, all the events from the `Helix' and `Violin mode' classes were not available, which is the reason our dataset has fewer classes than the Gravity Spy dataset.

The dataset is divided into training, testing and validation subsets, such that for each class $\sim 20\%$ of the spectrograms are used for testing, $\sim 10\%$ for validation and the rest are used for training. In total this results in training, test and validation sets consisting of 4249, 1197 and 482 spectrograms, respectively.

When training the network we use data augmentation to increase the size of the training set. 
The spectrogram images are randomly shifted by up to $50\%$ of the image size in the horizontal (time) direction (149 pixels, which are $1 \mathrm{s}$). Aside from increasing the size of the training set, which is necessary for training a deep network, the horizontal shift may be useful in training the network to identify spectrograms of signals which are not centered in the spectrogram (since we do not know in advance the timing of an event relative to the center of the spectrogram it appears in). 
To reduce class imbalance, for classes with more than 1000 spectrograms, we augment each spectrogram once (one spectrogram with the random shifts described above is created), for classes with between 200 and 1000 spectrograms we augment each spectrogram 4 times, and for classes with fewer than 200 spectrograms we augment each spectrogram 10 times. In total the augmented training set, including the original training set, consists of 24641 spectrograms.

\subsection{Training the network}
\label{ssec:train_net}

We train our CNN using the transfer learning method, used by \citet{george2017deep}.
We take a CNN that was pre-trained on ImageNet -- a large dataset of over 1 million images in 1000 different classes -- and fine-tune its weights by re-training it on the dataset described above. The pre-trained network we use is ResNet152V2 proposed by \cite{he2016identity}. 
This network is a refined version of the original residual network presented in \citet{he2016deep}, known as ResNet, which became one of the most popular architectures in the field of image processing using neural networks since it won the ILSVRC image classification competition in 2015.
The advantage of residual architectures is that they allow training very deep networks by addressing the vanishing gradient problem. The problem is that when using backpropagation to compute gradients in deep networks, the gradients of the initial layers become very small, effectively preventing training (see for instance \citep{glorot2010understanding}). 
The ResNet architecture deals with this problem by using `shortcut connections', which are connections that skip one or more layers. The basic building block of the ResNet architecture is a block of two or three convolution layers with ReLU activations and a `shortcut connection' connecting the input of the block to its output, before the output activation (skipping the convolution layers). The refinement presented in \citet{he2016identity} is to use `pre-activation' instead of `post-activation', meaning adding a ReLU activation layer at the input of the block and removing the one at its output. ResNet152V2 contains 50 such blocks containing 3 convolution layers each, plus an additional convolution layer at the network's input, and a fully connected layer of 1000 neurons with softmax activation at its output, totalling in 152 trainable layers.

\begin{figure}
	\includegraphics[width=\linewidth]{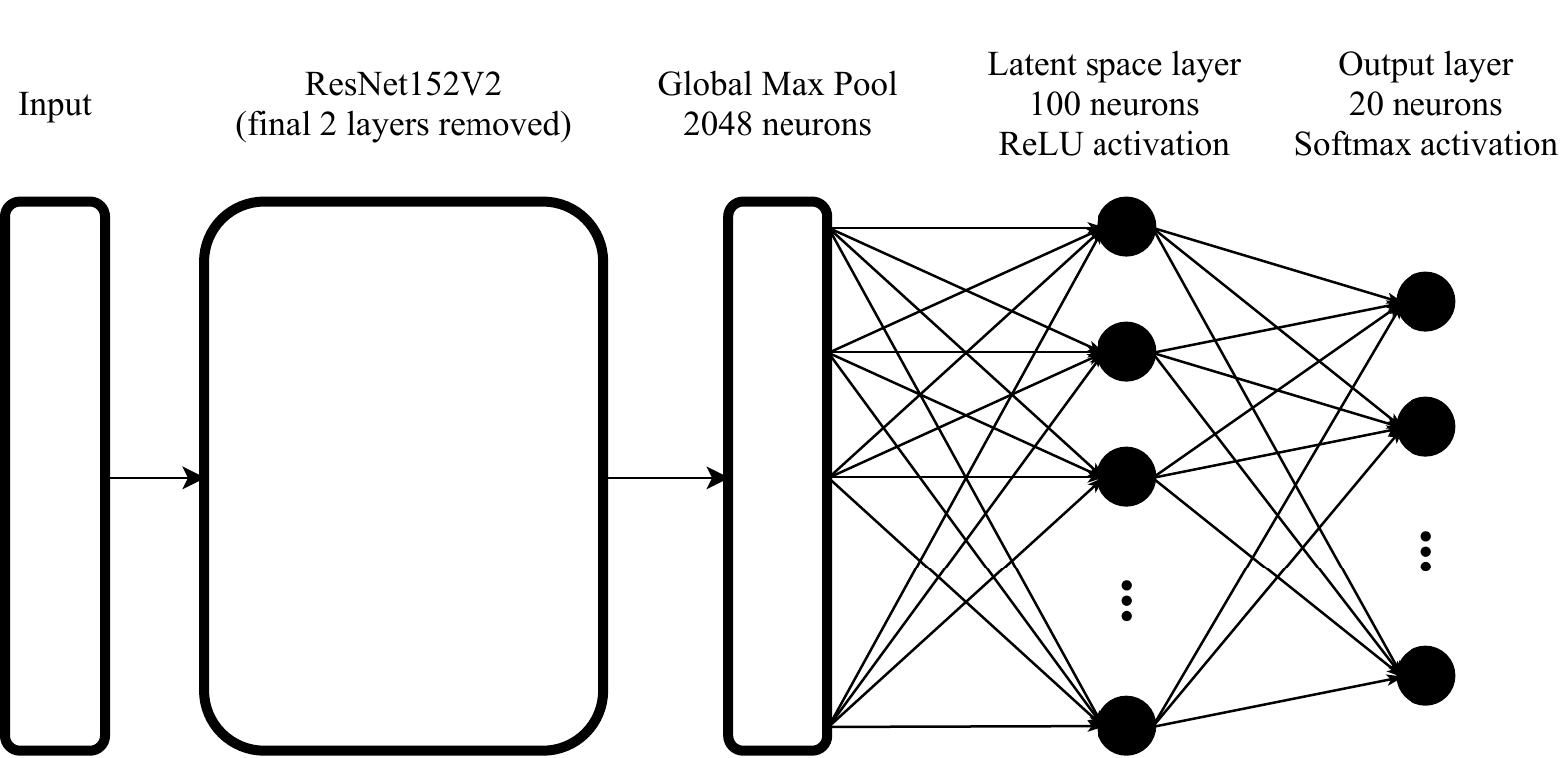}
	\caption{A block diagram of the network trained in this search. The input layer is the spectrogram data, which is fed to a a ResNet152V2 network pre-trained on the ImageNet dataset. The final layers of the pre-trained network are replaced with our classification network consisting of a global max pooling layer, followed by a fully connected layer with 100 neurons and ReLU activations and finally by a fully connected layer with 20 neurons and softmax activations.}
	\label{fig:cnn}
\end{figure}

We adapt ResNet152V2 for our task by removing the output layer as well as the global average pool layer just before it, and replacing them with a classification network for classifying the 20 classes in our dataset. The input layer of our classification network is a global max pool layer, the output layer is a fully connected layer with 20 neurons and  softmax activation, and we experimented with several architectures for the network layers in between. Ultimately the chosen architecture is a single fully connected layer with 100 neurons and ReLU activation between the input and output layers, since this seems to provide sufficient information with little redundancy. A summary of the architecture of our network is given in \cref{fig:cnn}.

We train the network described above for 20 epochs utilizing the cross entropy loss and ADADELTA optimization, using the \texttt{keras} python package. 
We choose the weights of the epoch with the highest classification accuracy on the validation set and use them throughout the rest of this work.

There are several differences in our application of transfer learning to GW data from what is described in \citet{george2017deep}. First, there are the differences in the dataset described in \cref{ssec:gen_spec,ssec:split}, 
the main being the fact that our dataset has two 
fewer classes. 
Second, we implement different data augmentation, allowing much larger horizontal shifts and removing vertical shifts and zoom augmentation used by \citet{george2017deep}. Third, we use ResNet152V2 as the pre-trained network, whereas \citet{george2017deep} experiment with several networks but not this particular one. In addition, we replace the global average pool layer with a global max pool layer and insert the additional 100 neuron fully connected layer. Finally, we do not add dropout layers to our network, and use ADADELTA optimization and not AdaGrad or ADAM since these resulted in degraded performance in our case.

The resulting network has a classification accuracy of $97.3\%$ on the test set, which is marginally lower than the accuracy achieved by \citet{george2017deep}. However, our goal is to use this network for an unsupervised GW signal search, not to classify glitches 
with the highest accuracy possible. Therefore, we are content with this result and did not take additional
steps to achieve higher classification accuracy.

\section{Outlier detection}
\label{sec:outlier_detection}

We test two outlier detection schemes which use the network described above. In the first method we use the network and a dimensionality reduction algorithm for feature extraction, estimate the probability distribution in the extracted feature space and define points in regions of low probability as outliers. We refer to this method as density based.
In the second method we use activations from layers throughout the network to directly define a deviation for a given spectrogram with respect to the training set. Specifically, this method characterizes activations using Gram matrices, therefore we refer to it as Gram matrix based.
The two methods are described in the following subsections.

\subsection{Density based method}
\label{ssec:density}

Each spectrogram in our dataset is fed through the network, and the latent layer just before the output layer is extracted as a  feature representation for the spectrogram in the 100D latent space. Next, we perform dimensionality reduction using the UMAP algorithm \citep{mcinnes2018umap}, 
which can compute a mapping between the 100D latent space and a lower dimensional space. We compute such a mapping to a 2D space (which we refer to as the map space) by training the UMAP algorithm on the latent features of the training examples in our dataset. The key in this step is to reduce the dimensionality of the data, to something manageable by a human, while preserving as much as possible the structure of the manifold in which the data reside. 

\begin{figure*}
	\includegraphics[width=\linewidth]{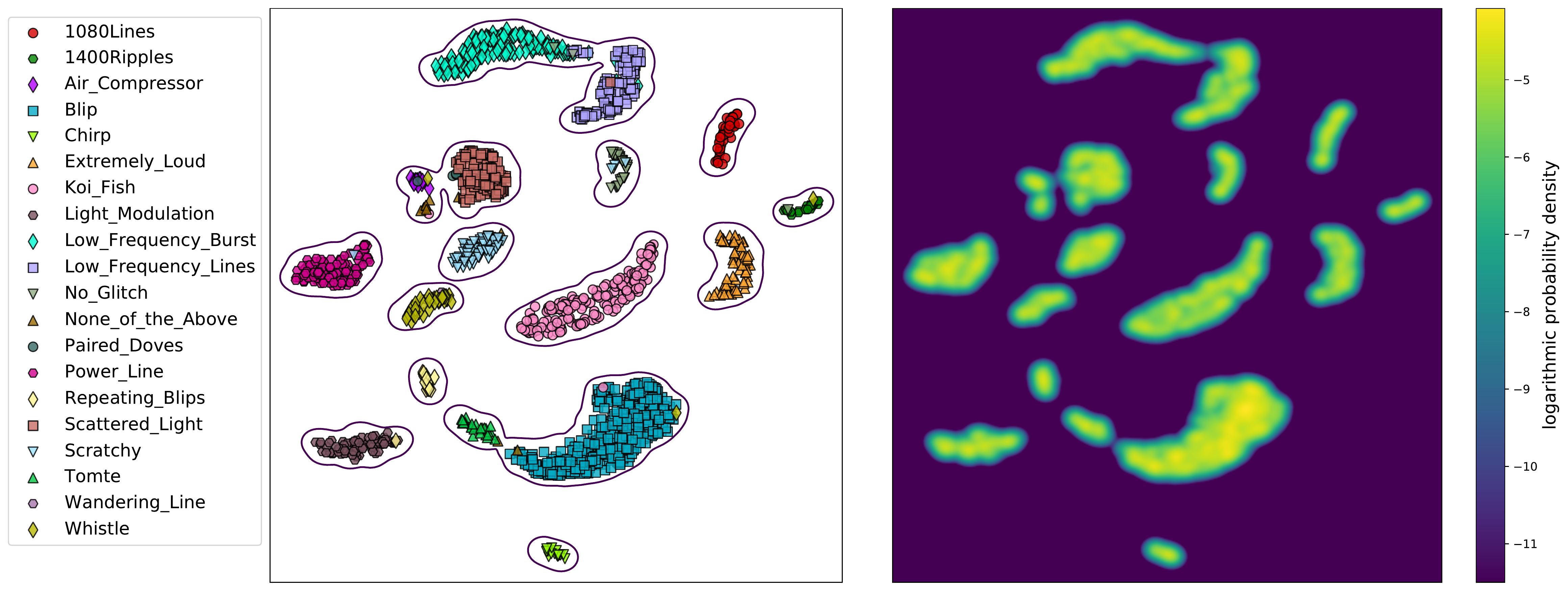}
	\caption{Plots illustrating the map space which is the result of training the UMAP algorithm on the latent space features of the training set. The axes of these plots are the two abstract dimensions of the map space. 
	The left panel is a plot of the map space representation of the spectrograms in the test and validation sets. Each point represents a spectrogram from the dataset, and their shapes and colors are determined according to the corresponding label. It can be seen that various classes cluster quite well. 
	The contours are contours of logarithmic probability density equal to $th=-11.5$ in this space. 
	The right panel is a heat map of the estimated distribution of points in the map space. The color corresponds to the logarithmic probability density.}
	\label{fig:umap}
\end{figure*}

There are many existing dimensionality reduction algorithms that one could choose from. The key advantages of UMAP are that it is fast to train, and that it can be trained on one dataset, and then used to transform new data without re-training the model.
We use this advantage in the search phase to map unlabeled spectrograms, and also as a qualitative indication of the robustness of the mapping, by mapping the latent features of the test and validation sets in our dataset. A visualization of the points representing the spectrograms in the test and validation sets is shown in the left panel of \cref{fig:umap}, and it can be seen that these points form relatively well-defined and separated clusters according to the different classes in the dataset.

The UMAP algorithm has a few hyper-parameters when training. We experimented with several values for \texttt{n\_neighbors} and \texttt{min\_dist} and ultimately chose the values 15 and 0.25 respectively, since these resulted in qualitatively good clusters (meaning relatively well-defined and separated clusters as mentioned above).

Now, we estimate the sample distribution of points in the map space using kernel density estimation. 
Each point in the sample is assigned a normalized kernel centered around it, and the kernels of all the points are summed and normalized to give the estimated distribution. We use the \texttt{scikit-learn} implementation to compute this estimated distribution, with a Gaussian kernel and a value of $0.3$ for the \texttt{bandwidth} parameter. The sample used to compute this estimation is the training set. Using this estimated distribution we can create a heat map of the map space, with the color corresponding to the logarithmic probability density of each point, which can be seen in the right panel of \cref{fig:umap}.

This probability estimation is used to define outliers in the spectrograms generated in the search phase by mapping them onto the map space in the same way as the spectrograms in our dataset, computing each resulting point's probability density and choosing a threshold below which the point is considered an outlier. The choice of the threshold determines how many points are flagged as outliers, therefore we choose it such that it is  feasible to go over all the flagged spectrograms manually. 
This is obviously somewhat arbitrary, and there is some degeneracy between the choice of the threshold and the kernel bandwidth when estimating density. We choose a threshold value of $th=-11.5$ which can be visualised by plotting contours of logarithmic probability density equal to this value in the map space, which are also shown in the left panel of \cref{fig:umap}. It can be seen that these contours surround the different clusters in the map space.

We note that the density estimation could have been calculated in the latent space, without applying dimensionality reduction, and outliers could be detected directly in this space. We experimented with this, but with no possibility to look at intermediate results in such space, we found the resulting outliers to be less promising than the ones detected in the map space.

\subsection{Gram matrix based method}
\label{ssec:gram}

This method was proposed by \citet{sastry2019detecting} in the context of out-of-distribution (OOD) detection by deep neural networks. The problem of detecting samples which are out of the distribution of the training set is especially important when deploying deep networks in real-world scenarios, since they can often output high confidence predictions for obviously OOD samples, including noise \citep{nguyen2015deep}. 
Because the network is trained to identify a finite set of classes, it attempts to match any input into one of these classes, even if it is clearly different. This is obviously a major limitation when using a classification CNN to search for outliers. The Gram matrix method attempts to remedy that, so we implement it in addition to the density method described in \cref{ssec:density}.
The main steps of the method are described below, and a more detailed description can be found in \citet{sastry2019detecting}.

We extract Gram matrices for a set of layers $L$ in our network, which are activation correlations between different channels within each layer. Explicitly, the Gram matrix of layer $l \in L$ is defined as:
\begin{equation}
	G_l = F_l F_l^T
\end{equation}
Where $F_l$ is the matrix representing the feature map of layer $l$ (meaning, the activations of the neurons in that layer). Its dimensions are $n_l \times p_l$ where $n_l$ and $p_l$ are the number of channels and number of neurons in each channel in this layer respectively. Since $G_l$ is symmetric, we extract the upper triangular matrix from it (with the diagonal) as a flattened array of $\frac{1}{2} n_l \left(n_l+1\right)$ elements, denoted by $\overline{G_l}$. For each class $C$ in the dataset, we compute the minimal and maximal values for the elements in these arrays over all the samples in that class, resulting in $mins_{C,l}$ and $maxs_{C,l}$ arrays, of $\frac{1}{2} n_l \left(n_l+1\right)$ elements each.
Now given a new sample $X$, for which the predicted class is $C$ we compute the layer-wise deviations from the minimal and maximal arrays for this class:
\begin{equation}
	\delta_l(X) = \sum_{i=1}^{\frac{1}{2} n_l \left(n_l+1\right)} {\delta\left(mins_{C,l}[i], maxs_{C,l}[i], \overline{G_l(X)}[i]\right)}
\end{equation}
Where
\begin{equation}
	\delta\left(min, max, val\right) = \begin{cases}
	0, & \text{if } min \leq val \leq max \\
	\frac{min - val}{\left|min\right|}, & \text{if } val < min \\
	\frac{val - max}{\left|max\right|}, & \text{if } val > max \\
\end{cases}
\end{equation}
Now we define the total deviation of the sample as the sum of the layer-wise deviations, normalized by the expected values of these deviations, computed using the validation set, $\mathbb{E}_{val}\left[\delta_l\right]$:
\begin{equation}
	\Delta(X) = \sum_{l \in L} {\frac{\delta_l(X)}{\mathbb{E}_{val}\left[\delta_l\right]}}
\end{equation}

The motivation for defining the deviation this way was to examine differences in the activity patterns of a sample throughout the network, relative to the patterns of the training examples in the class assigned to this sample. Activity patterns that significantly differ from those of the training examples in this class are a strong indication that this is an OOD sample. The Gram matrices are used to describe the activity patterns throughout the network since they were shown by \citet{gatys2016image} to be useful for encoding stylistic attributes of an image, such as textures and patterns.

Now we can define outliers in the spectrograms generated in the search phase by feeding each spectrogram through the network, computing its total deviation score and choosing a threshold above which a spectrogram is considered an outlier. We choose a threshold value of $th=5100$, in a similar way to the choice described in \cref{ssec:density}.

We note a few differences between our implementation of this method and the original proposed by \citet{sastry2019detecting}. First, \citet{sastry2019detecting} compute the Gram matrices and the layer-wise deviations for all the layers in their network. We compute these matrices and deviations for a subset of layers $L$. The layers we use are the output layers of the odd numbered residual blocks, 26 layers in total (not 25 because the layer numbering is not continuous). Second, \citet{sastry2019detecting} define higher-order Gram matrices, while we only use the first order matrix defined above. The reason for both differences is reduction in the time required to process each spectrogram, since the network we use is larger than the one used in \citet{sastry2019detecting} (they used ResNet34, which consists of 34 layers).

Due to the importance of OOD detection, there are quite a few more methods being proposed in the literature. The advantages of the one we chose to implement here are that it can be used with pre-trained networks, i.e., we can use the same network we trained in \cref{ssec:train_net}. Furthermore, it only has a few hyperparameters and can work without the need for fine-tuning using OOD examples, something we prefer not to bias ourselves with. In addition, this method is reported by \citet{sastry2019detecting} to have a performance comparable to the state-of-the-art.

\section{Search phase}
\label{sec:search}

\subsection{Generating spectrograms}
\label{ssec:search_gen}

We generate unlabeled spectrograms from bulk strain data using a process similar to the one described by \citet{robinet2016omicron}. 
First, the data are divided into segments for which data are continuously available (data are not publicly available for the entire duration of the observing runs), 
then we divide each segment into chunks of length $T_c = 64 \mathrm{s}$. Each chunk is conditioned by subtracting its mean, applying a high-pass filter with a cut-off frequency of $20 \mathrm{Hz}$, and finally it is multiplied by a Tukey window (using the \texttt{SciPy} implementation) which insures a smooth transition to 0 at both chunk ends, in order to avoid edge artifacts when generating the time-frequency spectrograms. In order to avoid data loss because of the windowing, there is an overlap of $T_o = 2 \mathrm{s}$ between consecutive chunks, which determines the parameter of the window to be $\alpha = T_o/T_c = 2/64$.

After conditioning, the data in each chunk are whitened: The noise power spectral density during the chunk is estimated and then a filter is applied to the chunk which yields a white (meaning constant) noise spectral density for the filtered data.

Next, the multi-Q transform, which is a modification of the constant-Q transform is applied to the whitened data. The constant-Q transform is a time-frequency transform similar to the standard short-time Fourier transform, which computes a time-dependant frequency representation of the data by computing windowed Fourier transforms with window lengths determining the frequency resolution, and the overlap between windows determining the time resolution. In the constant-Q transform the frequency axis is logarithmically spaced, and instead of using a constant window for all frequencies, the window length is inversely proportional to the frequency which results in a constant $Q$ value, which is the ratio of the frequency of each bin to its bandwidth. The multi-Q transform  applies multiple constant-Q transforms for a range of $Q$ values and then chooses the one which has highest peak energy. More detailed descriptions of these transforms and the way they are used for GW searches can be found in \citet{blankertz2001constant}, \citet{chatterji2004multiresolution} and \citet{robinet2016omicron}.

The multi-Q transform is computed using the \texttt{GWpy} method \texttt{q\_transform}, which also implements the data whitening. The function parameters we use in this work are the same as the ones that were used to generate the Gravity Spy dataset, except for the \texttt{gps} parameter. This parameter, when passed to the method, determines the time-stamp to focus on for choosing the $Q$ value with highest peak energy, and when the Gravity Spy spectrograms were generated this parameter was equal to the event time-stamp each spectrogram represents. We do not use this parameter, therefore the entire chunk is used for choosing the $Q$ value.

Finally, the resulting multi-Q transform is cropped by $T_o/2 = 1 \mathrm{s}$ at each end to get rid of data affected by the windowing, and the remaining $62 \mathrm{s}$ are divided into 31 non-overlapping spectrograms of $2 \mathrm{s}$ each.

The fact that our spectrograms do not overlap suggests we might miss signals that occur during the transition from one spectrogram to the next. One can overcome this issue by adding overlap between spectrograms, at a small computational cost. In this work we chose not to add this overlap, as the probability of missing signals is relatively small. The reason for this is that short signals have a small probability of occurring precisely during the transition between spectrograms, and we expect longer signals to create a significant pattern in at least one of the spectrograms even if they are truncated, such that they may still be detected as outliers by our search.

\subsection{Flagging spectrograms of interest}

We process the spectrograms using the network trained in \cref{ssec:train_net}, and flag outliers using the two methods described in \cref{sec:outlier_detection}. 

In addition to outlier detection, we also utilize the network predictions, and flag spectrograms that are classified as `Chirp' (the class of binary mergers). We also note that the `None of the Above' class is meant to be a `catch-all' class in the Gravity Spy dataset, containing all the glitches in the dataset that do not resemble any of the other classes. However, it would be false to assume that all outliers should be classified by the network as `None of the Above'. Rather, it will classify as `None of the Above' spectrograms which resemble the examples it was trained on, and there is no way of knowing in advance the prediction of an outlier with a spectrogram that is sufficiently different than all the spectrograms in the dataset (including the `None of the Above' examples). Indeed, when examining the spectrograms classified as `None of the Above' we found them less interesting than the outliers flagged by the methods described above. 
However, since the `None of the Above' should be a rare class, it is interesting to examine time-stamps for which the spectrograms from both detectors are classified as such.

\section{Results}
\label{sec:results}

We apply the methods described in \cref{sec:outlier_detection} to a subset of the public LIGO data from the first two observing runs -- O1 and O2 \citep{abbott2019open}. We focus on times where data are available from both LIGO detectors and process about 2 million seconds of data, which are about $13\%$ of those times. We generate spectrograms using the process described in \cref{ssec:search_gen} for the data from each detector, resulting in about 1 million spectrograms and corresponding 2D representations for each detector.

We examine the spectrograms flagged as outliers by the two methods, and divide them into 14 categories, summarized in \cref{tab:outliers} and described in detail in the following subsections.

\begin{table}
	\caption{Breakdown of the outliers detected by each method into different spectrogram categories.}
	\label{tab:outliers}
	\begin{tabular}{llcc}
		\hline
                                            &                               &	Density &	Gram matrix	\\
		\hline
		\multirow{11}{*}{Known glitches}    &   `$1080 \mathrm{Hz}$ lines'  &	2		&	2		    \\
                                            &   `Blip'-like					&	8		&	3			\\
                                            &   Loud glitch					&	2	    &	6			\\
                                            &   Low freq. patterns			&	77	    &	83			\\
                                            &   Multiple glitches			&	13		&	36		    \\
                                            &   `Power line'				&	7		&	0			\\
                                            &   `Scratchy'					&	2		&	2			\\
                                            &   Truncated glitch			&	41	    &	12			\\
                                            &   `Violin mode'				&	3	    &	2			\\
                                            &   `Wandering line'			&	3		&	13			\\
                                            &   `Whistle'					&	5		&	17			\\
        \hline
                                            &   Empty spectrogram			&	183		&	122			\\
                                            &   Hardware injection			&	0		&	25			\\
                                            &   Miscellaneous				&	40		&	55			\\
		\hline
                                            &   Total						&	386		&	378			\\
		\hline
	\end{tabular}
\end{table}

\subsection{Known glitches}
\label{ssec:glitches}

\begin{figure*}
	\includegraphics[width=\linewidth]{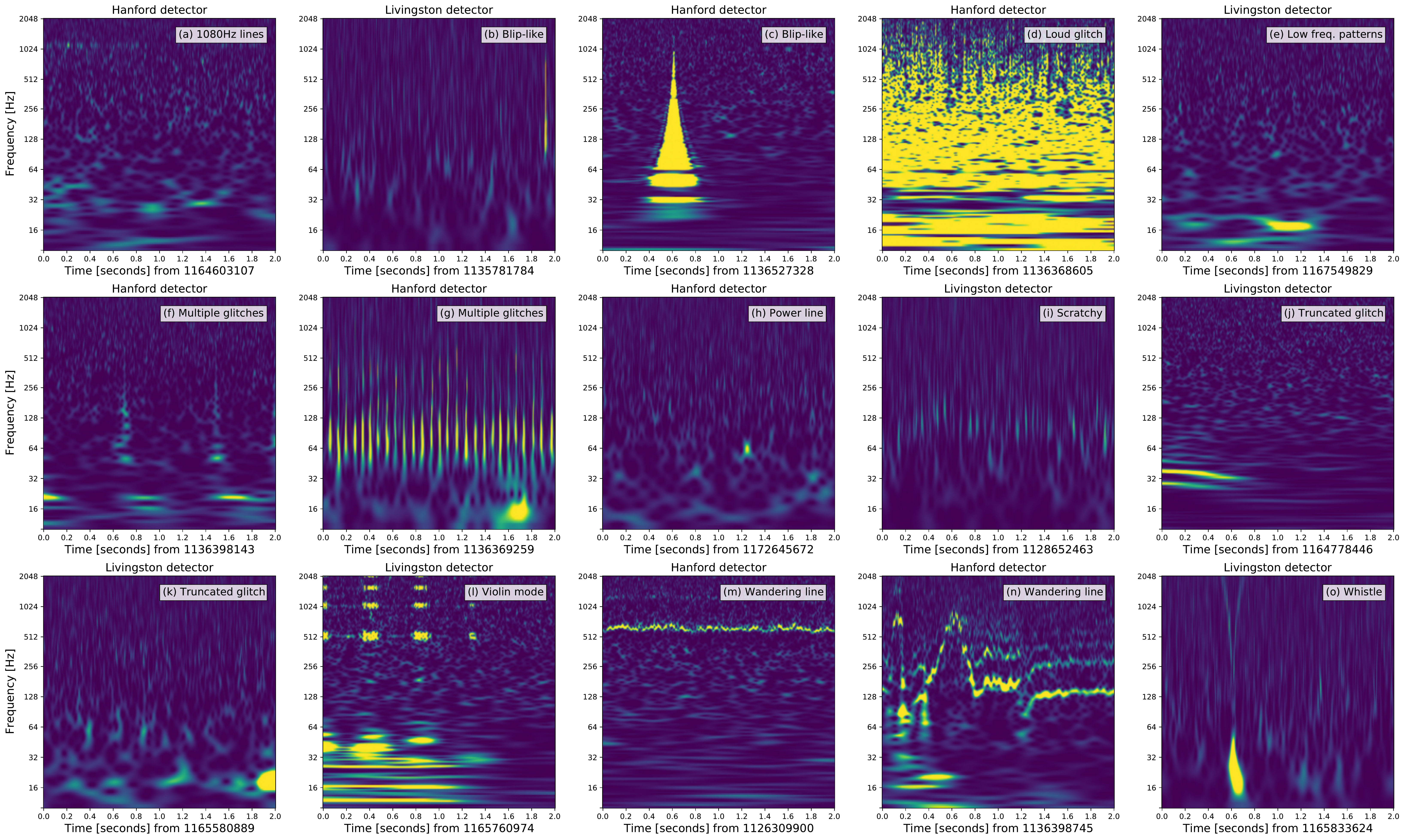}
	
	\phantomsubfloat{\label{fig:glitches_a}}
	\phantomsubfloat{\label{fig:glitches_b}}
	\phantomsubfloat{\label{fig:glitches_c}}
	\phantomsubfloat{\label{fig:glitches_d}}
	\phantomsubfloat{\label{fig:glitches_e}}
	\phantomsubfloat{\label{fig:glitches_f}}
	\phantomsubfloat{\label{fig:glitches_g}}
	\phantomsubfloat{\label{fig:glitches_h}}
	\phantomsubfloat{\label{fig:glitches_i}}
	\phantomsubfloat{\label{fig:glitches_j}}
	\phantomsubfloat{\label{fig:glitches_k}}
	\phantomsubfloat{\label{fig:glitches_l}}
	\phantomsubfloat{\label{fig:glitches_m}}
	\phantomsubfloat{\label{fig:glitches_n}}
	\phantomsubfloat{\label{fig:glitches_o}}
    \vspace{-2\baselineskip}
    
	\caption{Examples of outliers containing various known glitches. The x-axis in these (and subsequent) spectrogram plots is time, counted from the initial spectrogram time, stated in the x-axis label in GPS time.}
	\label{fig:glitches}
\end{figure*}

These spectrograms contain known types of glitches that are also present in the Gravity Spy dataset. These outliers, examples of which are shown in \cref{fig:glitches}, make up about half of the total outliers flagged by each method (163 and 176 by the density method and the Gram matrix method respectively), however, for many of them we can find a reasonable explanation for what might have caused them to be flagged as outliers. Many of them contain relatively faint signals, fainter than the typical glitches in the dataset. For instance, the spectrogram in \cref{fig:glitches_a} contains a faint `$1080 \mathrm{Hz}$ lines' glitch, the spectrogram in \cref{fig:glitches_h} contains a faint `Power line' glitch, and the spectrogram in \cref{fig:glitches_i} contains a faint `Scratchy' glitch. 
The fact these faint signals are flagged suggests that our search can detect these faint signals and identify that they are somewhat different from their louder counterparts, which is a desirable property of a search for unmodeled signals.
The spectrograms assigned to the multiple glitches category contain multiple glitches in a single spectrogram, like the ones shown in \cref{fig:glitches_f,fig:glitches_g}. The spectrograms assigned to the truncated glitch category contain glitches that overlap two consecutive spectrograms, and therefore the pattern in each one appears truncated, examples of which can be seen in \cref{fig:glitches_j,fig:glitches_k}. In addition, there are glitches which belong to classes which are under-represented in our dataset. For instance, the `Violin mode' glitch shown in \cref{fig:glitches_l} belongs to one of the classes mentioned in \cref{ssec:split} which are not present in our training dataset, and the `Wandering line' glitches shown in \cref{fig:glitches_m,fig:glitches_n} belong to the smallest class in our dataset (containing only 3 examples in the training set). Finally, the `Whistle' glitch shown in \cref{fig:glitches_o} presents a somewhat different pattern than the glitches belonging to the same class in our dataset -- the low frequency part of the signal is quite strong while the higher frequency pattern is much fainter.

\subsection{`Empty' spectrograms}

\begin{figure*}
	\includegraphics[width=\linewidth]{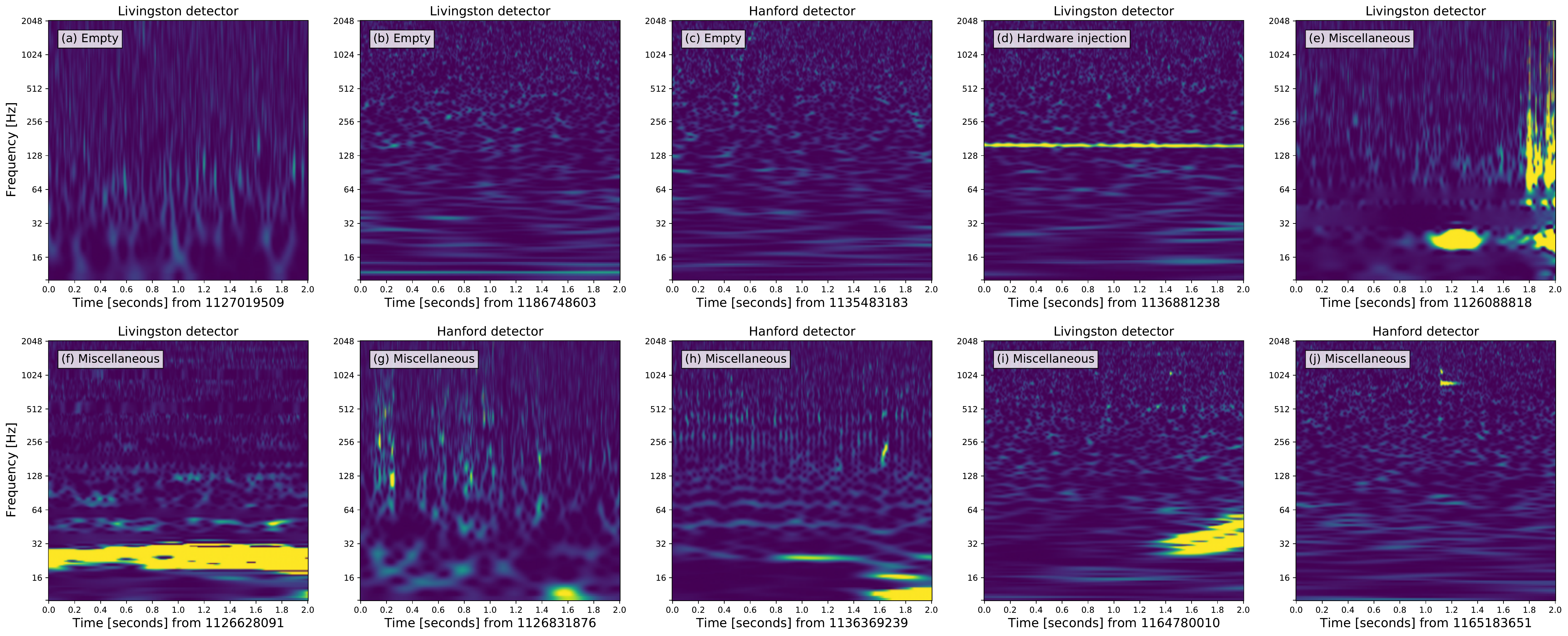}
	
	\phantomsubfloat{\label{fig:eim_a}}
	\phantomsubfloat{\label{fig:eim_b}}
	\phantomsubfloat{\label{fig:eim_c}}
	\phantomsubfloat{\label{fig:eim_d}}
	\phantomsubfloat{\label{fig:eim_e}}
	\phantomsubfloat{\label{fig:eim_f}}
	\phantomsubfloat{\label{fig:eim_g}}
	\phantomsubfloat{\label{fig:eim_h}}
	\phantomsubfloat{\label{fig:eim_i}}
	\phantomsubfloat{\label{fig:eim_j}}
    \vspace{-2\baselineskip}
    
	\caption{Examples of outliers assigned to the categories not associated with known glitches: Empty spectrograms, Hardware injections and Miscellaneous patterns.}
	\label{fig:empty_injection_misc}
\end{figure*}

These spectrograms appear to contain only background detector noise, with no noticeable distinct pattern. Currently we do not understand what caused these spectrograms to be flagged as outliers, however, the fact that several such spectrograms are also flagged by our search illustrates that not only very loud and obvious patterns are flagged, which is a requirement for being able to detect faint signals if they exist in the data. We note that it is not always clear visually whether a given spectrogram should be treated as `empty' or not. An illustration of this can be seen in the examples of `empty' spectrograms shown in \cref{fig:empty_injection_misc}. For instance, the spectrogram in \cref{fig:eim_b} might contain some low frequency signals and the one in \cref{fig:eim_c} might contain a faint `Whistle'-like signal. As mentioned in \cref{ssec:glitches} being able to flag such faint signals is a desirable property of this search. 

\subsection{Hardware injections}
\label{ssec:injections}

During the first two observing runs, simulated signals were injected into the GW detectors in order to validate the data analysis pipelines, as well as several other uses \citep{biwer2017validating}. 

The spectrogram shown in \cref{fig:eim_d} contains a hardware injection of a loud, long duration signal, which does not simulate any astrophysical signal, but is instead used for detector characterization. The spectrogram pattern created by this signal is significantly different from the patterns created by the classes in the dataset used for training our models, which is the reason it was flagged by our search as an outlier. 
All the hardware injections flagged by our search are loud, long duration signals, which create similar patterns to the example shown in \cref{fig:eim_d}.

\subsection{Miscellaneous spectrograms}

These spectrograms contain miscellaneous patterns that are sufficiently different from the spectrograms in the Gravity Spy dataset. Flagging such spectrograms is the goal of this search, however, all the spectrograms assigned to this category contain patterns that are only present in a single detector. They are therefore unlikely to be astrophysical. In addition, most can be easily interpreted as noise. For instance, the spectrograms shown in \cref{fig:eim_e,fig:eim_f} contain patterns in a wide range of frequencies that are probably  the tails of very loud glitches. The spectrogram in \cref{fig:eim_g} also contains wide-band patterns that are similar to the patterns created by wide-band random noise. The spectrogram in \cref{fig:eim_h} contains a low frequency glitch together with some faint line harmonics, and since the resulting pattern is very unusual it was assigned to the miscellaneous category. The spectrograms in \cref{fig:eim_i,fig:eim_i} contain peculiar patterns that do not appear in any of the other spectrograms flagged by our search. Since these patterns are relatively strong in these spectrograms, we expect that if their sources were of astrophysical origin, they would have made some visible signature in both detectors. Since each of them was only detected in a single detector we conclude that they are most likely not of astrophysical origin, but rather an unrecognized type of glitch. If indeed these are rare glitches, more will likely be found as more data is analysed.  

\subsection{Spectrograms classified as `Chirp'}

91 spectrograms are classified as `Chirp' by our network, from 74 distinct time-stamps (for 17 time-stamps the spectrograms from both detectors are classified as `Chirp').

\begin{figure*}
	\includegraphics[width=\linewidth]{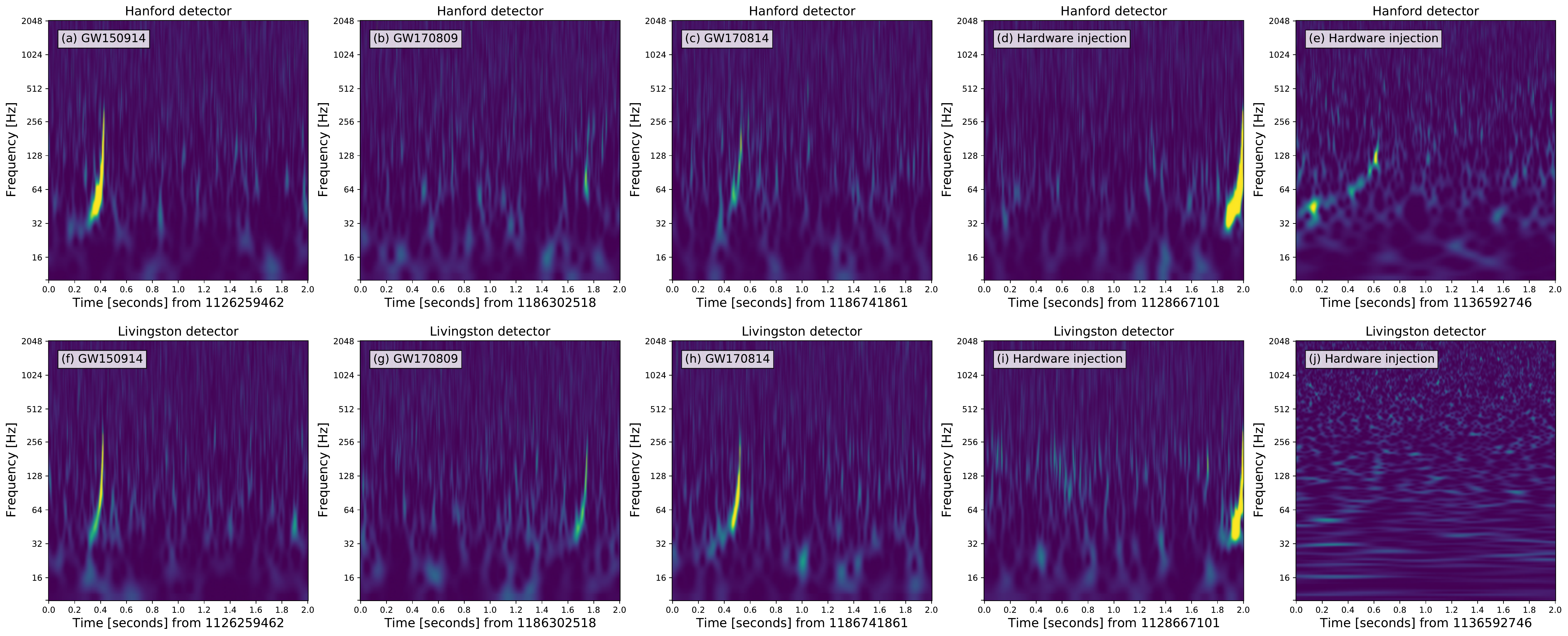}
	
	\phantomsubfloat{\label{fig:chirp_a}}
	\phantomsubfloat{\label{fig:chirp_b}}
	\phantomsubfloat{\label{fig:chirp_c}}
	\phantomsubfloat{\label{fig:chirp_d}}
	\phantomsubfloat{\label{fig:chirp_e}}
	\phantomsubfloat{\label{fig:chirp_f}}
	\phantomsubfloat{\label{fig:chirp_g}}
	\phantomsubfloat{\label{fig:chirp_h}}
	\phantomsubfloat{\label{fig:chirp_i}}
	\phantomsubfloat{\label{fig:chirp_j}}
    \vspace{-2\baselineskip}
    	
	\caption{Spectrograms from time-stamps classified as `Chirp'. In each column the top and bottom panels are the spectrograms from the Hanford and Livingston detectors respectively. The first three columns contain GW events detected in O1 and O2, and the final two columns contain hardware injections of simulated BH mergers.}
	\label{fig:chirp}
\end{figure*}

Three of these time-stamps contain GW events that were detected during O1 and O2, the spectrograms from these time-stamps can be seen in \cref{fig:chirp}. In each column the top and bottom panels show the spectrograms from the Hanford and Livingston detectors respectively, for the same time-stamp. For GW150914 (shown in \cref{fig:chirp_a,fig:chirp_f}) the spectrograms from both detectors were classified as `Chirp', while for the other two events (GW170809 and GW170814, shown in \cref{fig:chirp_b,fig:chirp_g} and \cref{fig:chirp_c,fig:chirp_h} respectively) only the spectrogram from Livingston was classified as `Chirp'.

22 of the time-stamps contain hardware injections of signals simulating BBH mergers, all but one injected into both detectors. For 16 of the 22 time-stamps the spectrograms from both detectors were classified as `Chirp'.
The example shown in \cref{fig:chirp_d,fig:chirp_i} contains a signal simulating the merger of two $38 \mathrm{M_\odot}$ BHs at a distance of $344 \mathrm{Mpc}$ which was injected into both detectors, and the spectrograms from both were classified as `Chirp'. This example is typical to most of the hardware injections classified as `Chirp' by our network -- in all but one, at least one BH has a mass over $30 \mathrm{M_\odot}$ and the other has a mass over $15 \mathrm{M_\odot}$, and the resulting spectrograms have fairly similar patterns. The one atypical injection is shown in \cref{fig:chirp_e,fig:chirp_j}. This injection is of a signal simulating the merger of two BHs of masses $23 \mathrm{M_\odot}$ and $6 \mathrm{M_\odot}$ at a distance of $433 \mathrm{Mpc}$, which was also injected into both detectors. It is interesting to note that this signal is a much longer signal than the other injections ($\sim 6 \mathrm{s}$ in comparison to $<0.5 \mathrm{s}$), but the spectrogram that was classified as `Chirp' contains only the final $\sim 0.6 \mathrm{s}$ of the signal. However, only the spectrogram from Hanford was classified as `Chirp'.

26 of the spectrograms classified as `Chirp' contain `Blip'-like glitches, which can sometimes resemble very short duration chirps. All of these appear as a fairly loud signal in a single detector, therefore, we do not suspect they contain an astrophysical signal that was not detected previously, but treat them as glitches that were misclassified as `Chirp'.

The remaining 23 spectrograms contain miscellaneous patterns and we do not know why they were classified as `Chirp', but again, their loudness and appearance in a single detector indicate that they are noise patterns that were misclassified.

While the goal of this work is not to search for binary mergers using a classification neural network, and indeed we do not claim to have better sensitivity than matched-filtering, the spectrograms in this section are presented as further support for the utility of using a neural network for processing GW data. Both the the true positive `Chirps' (GW events and hardware injections) as well as the small number of false positives (49 out of about 1 million time-stamps) suggest high performance, as also seen by \citet{george2018deep,gabbard2018matching,krastev2020real} and \citet{schafer2020detection}.

\subsection{Spectrograms classified as `None of the Above'}

796 spectrograms are classified as `None of the Above', however, in all of the corresponding time-stamps there is no coincidence in the other detector, therefore we do not discuss them further. The fact that we did not find a single candidate for which both spectrograms are classified as `None of the Above' further suggests that these are rare and worthy of scrutiny, if and when they occur.

\section{Evaluating the search}
\label{sec:eval}

As naturally is the case with unsupervised or semi-supervised tasks, it is hard to assess the quality and robustness of the results. Did we not find any interesting source because of weaknesses in our methods, or because there are none in the data? We attempt to examine that question via numerical simulations across the vast parameter space of possible waveforms.  
We do so by injecting simulated signals to the data and examining the outcome of the search for these signals.

\subsection{Ad-hoc waveforms}
\label{ssec:adhoc}

We inject signals with a variety of ad-hoc waveforms similarly to what is described in \citet{abadie2012all} and \citet{abbott2017all}. First, we inject each of the generic burst waveforms described in \citet{abadie2012all} varying the relevant parameters and a with range of amplitudes around the $h_{rss}$ values at which \citet{abadie2012all} and \citet{abbott2017all} reported to achieve $50\%$ detection efficiency at a false alarm rate (FAR) of 1 in 100 yr. 
$h_{rss}$ is the root-sum-square strain amplitude of the signal, defined as:
\begin{equation}
	h_{rss} = \sqrt{\int{ \left({\left \vert h_+ (t) \right \vert}^2 + {\left \vert h_\times (t) \right \vert}^2 \right) dt}}
\end{equation}
where $h_+ (t)$ and $h_\times (t)$ are the plus and cross polarizations of the gravitational waveform respectively. The waveforms we inject are elliptically polarized with an ellipticity $\alpha$ (defined explicitly in \citealt{abadie2012all}) uniformly distributed in the interval $\left[0, 1\right]$.
Each of these waveforms is injected into both detectors at 15 random time-stamps which contain only background detector noise, using a random, isotropically distributed sky position $\left(\Theta, \Phi \right)$ and a uniformly distributed polarization angle $\Psi$ in order to compute the antenna patterns of the detector $F_+$ and $F_\times$. The resulting strain at the detector is given by: 
\begin{equation}
	h_{det} \left(t \right) = F_+ \left(\Theta, \Phi, \Psi \right) h_+\left(t \right) + F_\times \left(\Theta, \Phi, \Psi \right) h_\times\left(t \right)
\end{equation}
\Cref{tab:generic_injections} summarizes the parameters and amplitude ranges of the generic burst waveforms we injected.

\begin{table}
	\caption{Summary of the injection parameters used when injecting the generic burst waveforms, and the $h_{rss}$ range for each, in units of $10^{-22} \mathrm{{Hz}^{-1/2}}$.}
	\label{tab:generic_injections}
	\begin{tabular}{lc}
		\hline
		Waveform		&	$h_{rss}$ range	\\
		\hline
		\\[\dimexpr-\normalbaselineskip+.1cm]
		Gaussian							&						\\
		$\tau = 0.1 \mathrm{ms}$	&	$30-50$		\\
		$\tau = 2.5 \mathrm{ms}$	&	$30-50$		\\[.1cm]
		
		Sine-Gaussian										&					\\
		$f_0 = 70 \mathrm{Hz}$, $Q = 100$		&	$5-30$		\\
		$f_0 = 235 \mathrm{Hz}$, $Q = 100$	&	$5-30$		\\
		$f_0 = 554 \mathrm{Hz}$, $Q = 9$		&	$5-30$		\\
		$f_0 = 849 \mathrm{Hz}$, $Q = 3$		&	$5-30$		\\
		$f_0 = 1614 \mathrm{Hz}$, $Q = 100$	&	$5-30$		\\
		$f_0 = 2000 \mathrm{Hz}$, $Q = 3$		&	$5-30$		\\[.1cm]
		
		Ring-down																	&						\\
		$f_0 = 2000 \mathrm{Hz}$, $\tau = 1 \mathrm{ms}$		&	$30-50$		\\
		$f_0 = 2000 \mathrm{Hz}$, $\tau = 0.2 \mathrm{s}$	&	$30-50$		\\[.1cm]
		
		White-noise																			&					\\
		$f_{low} = 100 \mathrm{Hz}$, $\Delta f = 100 \mathrm{Hz}$, 
		$\tau = 0.1 \mathrm{s}$														&	$5-30$		\\
		$f_{low} = 250 \mathrm{Hz}$, $\Delta f = 100 \mathrm{Hz}$, 
		$\tau = 0.1 \mathrm{s}$														&	$5-30$		\\
		\hline
	\end{tabular}
\end{table}

Both outlier detection methods described in \cref{sec:outlier_detection} fail to flag these generic burst waveform injections as outliers. However, we note that these waveforms generally create short, quite simple spectrogram patterns, and that some of them are very similar to some of the glitches which our search is trained to identify as inliers (mostly the `Blip'-like glitches -- the `Blip', `Tomte' and `Koi-fish' glitch classes). 

We inject additional signals which create more intricate spectrogram patterns (mainly, the signal frequency has some evolution throughout the duration of the spectrogram). The additional waveforms are sweeps of linearly decreasing frequency, sweeps of linearly increasing frequency and the sum of two linearly increasing sweeps, all multiplied by a Gaussian envelope. While we do not have a physical justification for the linearly decreasing sweeps, some motivation for including the linearly increasing sweeps is the fact that several simulations for the GW signal produced by core-collapse supernovae exhibit some component of linearly increasing frequency (for instance \citealt{mezzacappa2020gravitational,zha2020gravitational}). For all sweeps we generate two polarizations by choosing the initial phase of the sweep, where the plus and cross polarizations have initial phases of $0$ and $\pi/2$ respectively, similarly to the sine-Gaussian and ring-down waveforms \citep{abadie2012all}.

The parameters defining the sweeps are $f_0$ - the initial frequency, $Q$ - the quality factor controlling the width of the Gaussian envelope (same as the sine-Gaussian waveforms), $t_1$ - the time at which $f_1$ is specified, and $f_1$ - the frequency at $t_1$. The double sweep waveforms are constructed by summing two sweeps, both starting from the same $f_0$, therefore there is an additional frequency $f_2$ - the frequency of the second sweep at $t_1$, and $a$ - the relative amplitude between the two signals. The decreasing sweeps are injected with a set of initial frequencies: $f_0 = 30, 70, 235, 849, 1615, 2000 \mathrm{Hz}$, each with four different quality factors: $Q = 3, 9, 100, 1000$, and $f_1 = f_0 / 2$. The parameters of the increasing and double sweeps are presented in \cref{tab:sweep_injections}. $f_0$, $Q$, and $f_1$ are the same for both waveform families, and $f_2$ and $a$ apply for the double sweep injections. For all sweeps we choose $t_1$ to be $t_1 = \tau = Q/ \left(\sqrt{2} \pi f_0\right)$. Each waveform is injected with four $h_{rss}$ values: $h_{rss} = \{5, 10, 20, 30\} \cdot 10^{-22} \mathrm{{Hz}^{-1/2}}$. Initially, each set of parameters is injected into 15 time-stamps, similarly to the generic burst waveforms. 

\begin{table}
	\caption{Increasing and double sweep injection parameters.}
	\label{tab:sweep_injections}
	\begin{tabular}{lcccc}
		\hline
		$f_0 \mathrm{[Hz]}$	&	$Q$			&	$f_1 \mathrm{[Hz]}$	&	$f_2 \mathrm{[Hz]}$	&	$a$	\\
		\hline
		30							&	3, 9, 100	&	600							&	60							&	0.6	\\
		70								&	9, 100		&	1400							&	84							&	0.2	\\
		\hline
	\end{tabular}
\end{table}

\begin{figure*}
	\includegraphics[width=.8\linewidth]{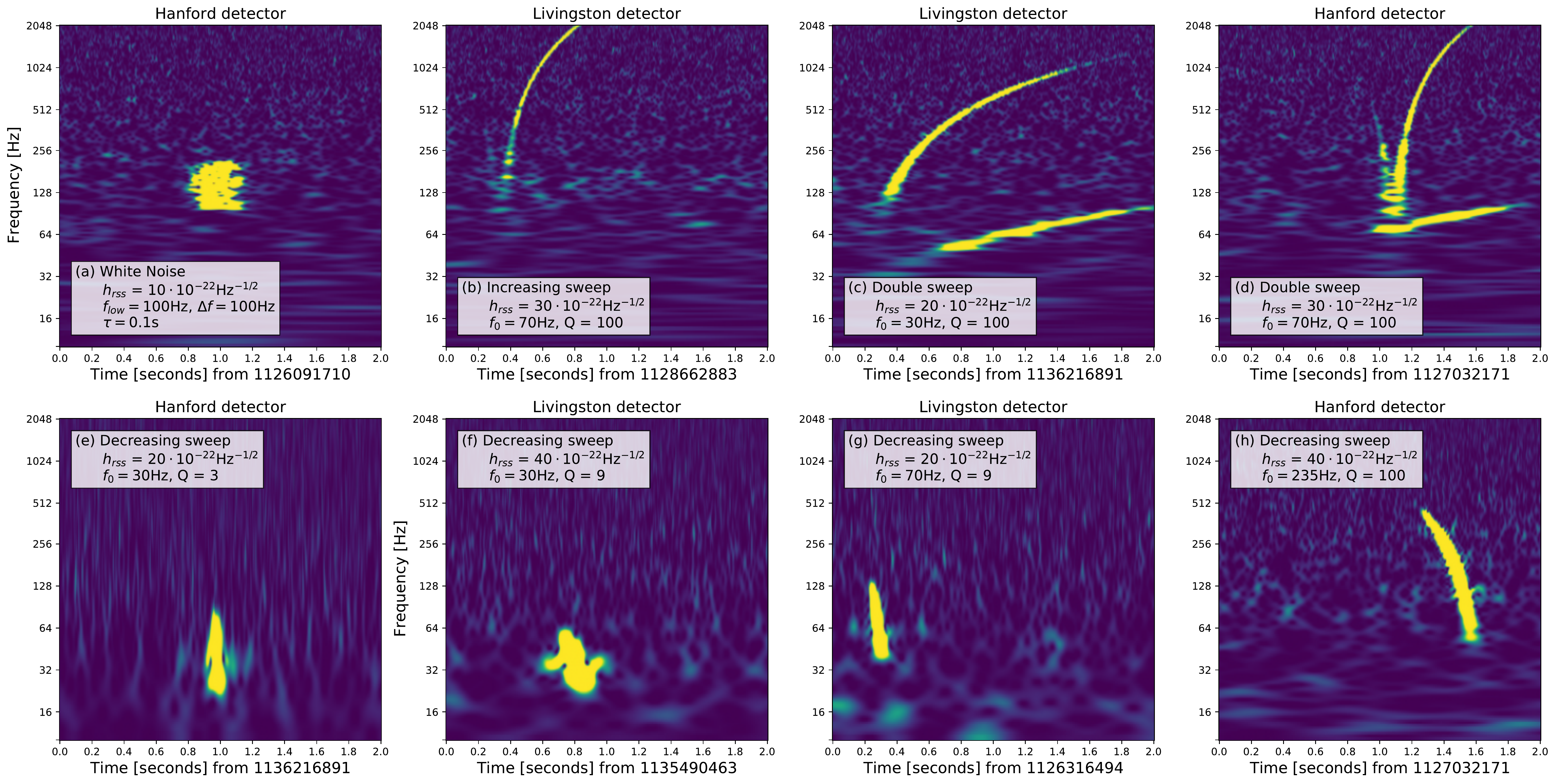}
	
	\phantomsubfloat{\label{fig:inj_a}}
	\phantomsubfloat{\label{fig:inj_b}}
	\phantomsubfloat{\label{fig:inj_c}}
	\phantomsubfloat{\label{fig:inj_d}}
	\phantomsubfloat{\label{fig:inj_e}}
	\phantomsubfloat{\label{fig:inj_f}}
	\phantomsubfloat{\label{fig:inj_g}}
	\phantomsubfloat{\label{fig:inj_h}}
    \vspace{-2\baselineskip}
    	
	\caption{Examples of injection spectrograms that are flagged as outliers by our search in both detectors.}
	\label{fig:inj}
\end{figure*}

The density based outlier detection method fails to flag these injections as well, however the Gram matrix method does flag some of the increasing sweeps and the double sweeps. 
In addition, some of the decreasing sweeps, as well as some of the white-noise injections described in \cref{tab:generic_injections} are classified as `None of the Above' in both detectors.
In order to gather additional statistics, the waveforms for which over $\sim 50\%$ of the injections are flagged in both detectors are injected into 1000 additional time-stamps, this time chosen at random without the constraint that they contain only background noise. \Cref{tab:detection_statistics} shows the waveform parameters and minimal amplitudes for which over $50\%$ of injections are flagged. In \cref{fig:inj} we show examples of spectrograms of these injections (for each injection we show the spectrogram from a single detector, since the patterns created by the injection are relatively similar in both detectors).

\begin{table}
	\caption{Waveform parameters for which $50\%$ of injections are flagged in both detectors. As before, the $h_{rss}$ values are presented in units of $10^{-22} \mathrm{{Hz}^{-1/2}}$. The method indicates the method by which the spectrograms are flagged -- either the Gram matrix outlier detection method (Gram), or the spectrograms from both detectors are classified as `None of the Above' (NotA).}
	\label{tab:detection_statistics}
	\begin{tabular}{lcc}
		\hline
		Waveform																			&	$h_{rss}$	&	Method	\\
		\hline
		\\[\dimexpr-\normalbaselineskip+.1cm]
		\multicolumn{3}{l}{White Noise}																				\\
		$f_{low} = 100 \mathrm{Hz}, \Delta f = 100 \mathrm{Hz}, 
		\tau=0.1 \mathrm{s}$														&	10				&	NotA		\\[.1cm]
		
		\multicolumn{3}{l}{Increasing sweep}																	\\
		$f_0 = 70 \mathrm{Hz}, Q = 100$										&	30			&	Gram	\\[.1cm]
		
		\multicolumn{3}{l}{Double sweep}																			\\
		$f_0 = 30 \mathrm{Hz}, Q = 100$										&	20			&	Gram 	\\
		$f_0 = 70 \mathrm{Hz}, Q = 100$										&	30			&	Gram 	\\[.1cm]
		
		\multicolumn{3}{l}{Decreasing sweep}																	\\
		$f_0 = 30 \mathrm{Hz}, Q = 3$											&	20			&	NotA		\\
		$f_0 = 30 \mathrm{Hz}, Q = 9$											&	40			&	NotA		\\
		$f_0 = 70 \mathrm{Hz}, Q = 9$											&	20			&	NotA		\\
		$f_0 = 235 \mathrm{Hz}, Q = 100$									&	40			&	NotA		\\
		\hline
	\end{tabular}
\end{table}

We note that only the increasing and double sweeps with a quality factor of $Q = 100$ are robustly flagged by the Gram matrix method. The lower $Q$ factor signals of all three sweep families correspond to shorter signals, which can also somewhat resemble the `Blip'-like glitches, similarly to the generic bursts mentioned above. In addition, both the increasing and decreasing sweeps can resemble the `Whistle' glitch class, which causes some of the higher $Q$ factor signals to not be flagged either.
These signals can be thought of as near-distribution samples -- outliers sampled from a distribution near the inlier distribution. Since the distribution is near the inlier distribution, 
combined with the fact the network attempts to fit any input into one of the classes it was trained on (as mentioned in \cref{ssec:gram}),
the samples are mapped to high density regions in the map space, which is the reason the density method does not flag them. The Gram matrix method is also reported to have decreased performance for near-distribution outliers \citep{sastry2019detecting}.
Nevertheless, some of the decreasing sweeps are robustly classified by the network as `None of the Above', mostly at low $Q$ factors, since these create patterns that are similar to some of the `None of the Above' spectrograms in the training dataset.

Although most of the signals injected are not robustly detected, we did manage to generate outliers that are flagged as such, even at relatively low amplitudes, which is additional support that if a sufficiently unusual signal exists it might be detected using our search. Our search is clearly far from optimal, but improvement should be reachable in future works. 

\subsection{Simulated CCSNe waveforms}

In addition to the ad-hoc waveforms described in \cref{ssec:adhoc}, we also inject simulated CCSNe waveforms taken from several studies \citep{abdikamalov2014measuring,andresen2019gravitational,radice2019characterizing}\footnote{\href{https://sntheory.org/ccdiffrot}{https://sntheory.org/ccdiffrot}}\textsuperscript{,}\footnote{\href{https://wwwmpa.mpa-garching.mpg.de/ccsnarchive/data/Andresen2019/}{https://wwwmpa.mpa-garching.mpg.de/ccsnarchive/data/Andresen2019/}}\textsuperscript{,}\footnote{\href{https://www.astro.princeton.edu/~burrows/gw.3d/}{https://www.astro.princeton.edu/~burrows/gw.3d/}}. 
Since these simulated waveforms are generated with different sampling rates and durations, they are first prepared for injection similarly to the process described by \citet{chan2020detection}. We resample the waveforms to the sample rate of the public detector data we inject them into, and then apply a high-pass filter with a cut-off frequency of $11 \mathrm{Hz}$ and a Tukey window with a parameter of $\alpha = 0.08$ in order to reduce artifacts. We rescale each waveform to several amplitudes to simulate progenitor distances between $0.2$ and $10 \mathrm{kpc}$, and inject each one using the same injection scheme used for the ad-hoc waveforms.

\begin{figure*}
	\includegraphics[width=\linewidth]{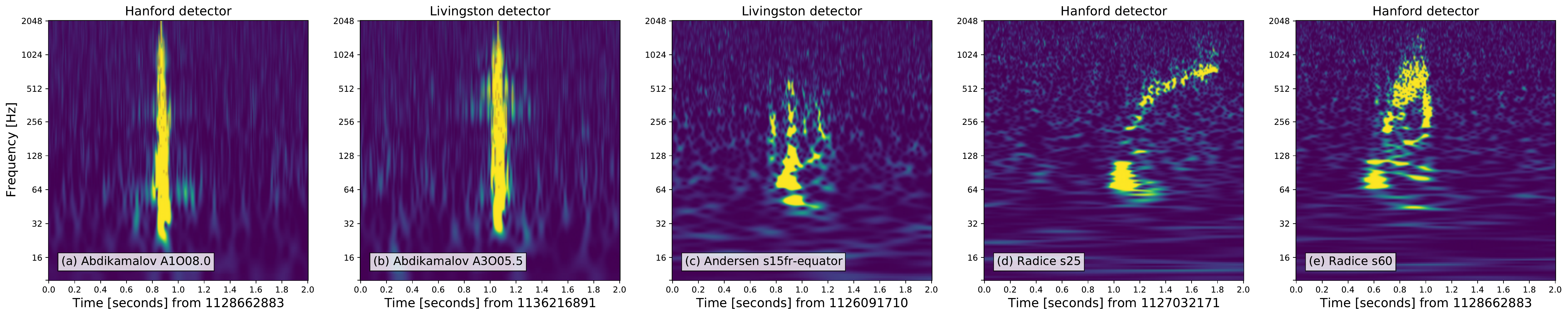}
	
	\phantomsubfloat{\label{fig:ccsn_inj_a}}
	\phantomsubfloat{\label{fig:ccsn_inj_b}}
	\phantomsubfloat{\label{fig:ccsn_inj_c}}
	\phantomsubfloat{\label{fig:ccsn_inj_d}}
	\phantomsubfloat{\label{fig:ccsn_inj_e}}
    \vspace{-2\baselineskip}
    	
	\caption{Examples of simulated CCSNe waveform spectrograms that are flagged as outliers by our search in both detectors.}
	\label{fig:ccsn_inj}
\end{figure*}

Both outlier detection algorithms fail to detect most of the injected waveforms at all simulated distances (only at $0.2 \mathrm{kpc}$ a small number of them are flagged). For some of the waveforms however, over $50\%$ of injections placed at $0.2 \mathrm{kpc}$ are classified as `None of the Above' in both detectors. These waveforms are presented in \cref{tab:ccsn_detection_statistics}, and examples of their spectrograms are shown in \cref{fig:ccsn_inj}.

\begin{table}
	\caption{Simulated CCSNe waveforms for which over $50\%$ of injections placed at $0.2 \mathrm{kpc}$ are flagged as `None of the Above' in both detectors. The $h_{rss}$ value denotes the amplitude of the signal at the detectors, presented in the same units as the previous tables.}
	\label{tab:ccsn_detection_statistics}
	\begin{tabular}{lcc}
		\hline
		Waveform												&	Progenitor mass $\mathrm{[M_\odot]}$		&	$h_{rss}$	\\
		\hline
		\\[\dimexpr-\normalbaselineskip+.1cm]
		\multicolumn{2}{l}{\citet{abdikamalov2014measuring}}																	\\
		A1O6.5													&	12																&	141			\\
		A1O7.5													&	12																&	172			\\
		A1O8.0													&	12																&	177			\\
		A1O9.0													&	12																&	212			\\
		A2O6.0													&	12																&	180			\\
		A2O6.5													&	12																&	172			\\
		A2O7.0													&	12																&	176			\\
		A3O5.0													&	12																&	150			\\
		A3O5.5													&	12																&	151			\\[.1cm]
		
		\multicolumn{2}{l}{\citet{andresen2019gravitational}}																	\\
		s15fr-equator										&	15																&	13				\\
		s15fr-pole												&	15																&	19				\\[.1cm]
		
		\multicolumn{2}{l}{\citet{radice2019characterizing}}																		\\
		s19														&	19																&	39			\\
		s25														&	25																&	29			\\
		s60														&	60								    							&	18			\\
		\hline
	\end{tabular}
\end{table}

There are several explanations why our search fails to detect waveforms at distances farther than $0.2 \mathrm{kpc}$. The waveforms from \citet{abdikamalov2014measuring} have short durations and aren't detected for the same reason as the short duration ad-hoc waveforms, as discussed in \cref{ssec:adhoc}. Indeed at $0.5 \mathrm{kpc}$ and above, most of the waveforms are classified by the network as either `Blip', `Koi-Fish' or `No-Glitch' when the signal amplitude is too low to create noticeable spectrogram patterns. The waveforms from \citet{andresen2019gravitational} and \citet{radice2019characterizing} have lower amplitudes to begin with, and therefore already at $0.5 \mathrm{kpc}$ a significant portion of them are classified as `No-Glitch', and at $3 \mathrm{kpc}$ and above all of them are classified as such.

While this is obviously not sensitive enough to be of practical use, the waveforms that are detected by our search  illustrate the capability of the search to detect signals that generate more complex spectrogram patterns, as seen by \cref{fig:ccsn_inj_c,fig:ccsn_inj_d,fig:ccsn_inj_e}.

\section{Conclusions}
\label{sec:conlustions}

In this paper we presented a novel approach to unmodeled GW searches using ML. This approach combines a supervised CNN, which we used to `learn' characteristics of GW strain spectrograms, with outlier detection algorithms used to flag spectrograms containing unusual patterns.

We ran a search over a significant part of the public LIGO data, about 13\% of the conincident stream from O1 and O2.  While we found no promising astrophysical signals in this search, the spectrograms that were flagged show the potential of such an approach.

The simulated signals that were detected by the search provide further support for using methods such as these, and the ones that were not detected present some limitations of the current implementation, which we will strive to improve in the future. In particular, our search did not detect most of the shorter duration signals (which have generally been the focus of previous unmodeled searches, \citealt{abadie2012all,abbott2017all}). Using shorter duration spectrograms might improve the sensitivity of detecting short signals (our choice of using $2 \mathrm{s}$ spectrograms was made early on but can be changed in future implementations). 

Another element that is worth further research is the CNN used for processing the spectrograms. The CNN architecture we used was designed to process `real-world' images, which have different properties than GW spectrograms (for instance, the spectrogram axes have different meanings, and a rotated spectrogram pattern has a different interpretation than a rotated image). While this architecture was particularly chosen to utilize the high performance image recognition networks available, a network designed specifically to process spectrograms might yield better results. More training data, or an improved training technique that utilizes the large amount of unlabeled data available, similar to the one described by \citet{noroozi2017seven}, should also improve the results. These and other possible improvements to the current implementation suggest there is still untapped potential in the approach presented here.

\section*{Acknowledgements}

This research has made use of the following \texttt{Python} packages: 
\texttt{IPython} \citep{perez2007ipython}, 
\texttt{gwPy} \citep{macleod2020gwpy}, 
\texttt{holoviews} \citep{philipp_rudiger_2019_3551257}, 
\texttt{keras} \citep{chollet2015keras}, 
\texttt{Matplotlib} \citep{hunter2007matplotlib}, 
\texttt{NumPy} \citep{harris2020array}, 
\texttt{pandas} \citep{mckinney2010data}, 
\texttt{PyCBC} \citep{alex_nitz_2020_3993665}, 
\texttt{scikit-image} \citep{van2014scikit}, 
\texttt{scikit-learn}, \citep{pedregosa2011scikit}, 
\texttt{SciPy} \citep{virtanen2020scipy} and 
\texttt{UMAP} \citep{mcinnes2018umap}.

This research has made use of data, software and/or web tools obtained from the Gravitational Wave Open Science Center (\href{https://www.gw-openscience.org}{https://www.gw-openscience.org}), a service of LIGO Laboratory, the LIGO Scientific Collaboration and the Virgo Collaboration. LIGO is funded by the U.S. National Science Foundation. Virgo is funded by the French Centre National de Recherche Scientifique (CNRS), the Italian Istituto Nazionale della Fisica Nucleare (INFN) and the Dutch Nikhef, with contributions by Polish and Hungarian institutes.

This research was supported by Grant No 2018017 from the United States-Israel Binational Science Foundation (BSF).
D.P. and T.M. acknowledge support from  Israel Science Foundation (ISF) grant 541/17.

\section*{Data Availability}

The data underlying this article are available in the Gravitational Wave Open Science Center, at \href{https://www.gw-openscience.org}{https://www.gw-openscience.org}.

\bibliographystyle{mnras}
\bibliography{gwml}


\bsp	
\label{lastpage}
\end{document}